# Reversible A ↔ B reaction – diffusion process with initially mixed reactants: boundary layer function approach

M. Sinder

*Department of Materials Engineering, Ben - Gurion University of the Negev, P.O.Box 653, Beer Sheva, 84105, Israel;*

V. Sokolovsky

*Physics Department, Ben - Gurion University of the Negev, P.O.Box 653, Beer Sheva, 84105, Israel*

J. Pelleg

*Department of Materials Engineering, Ben - Gurion University of the Negev, P.O.Box 653, Beer Sheva, 84105, Israel*

**Abstract**

The reversible A ↔ B reaction-diffusion process, when species A and B are initially mixed and diffuse with different diffusion coefficients, is investigated using the boundary layer function method. It is assumed that the ratio of the characteristic time of the reaction to the characteristic time of diffusion is taken as a small parameter of the task. It was shown that diffusion-reaction process can be considered as a quasi-equilibrium process. Despite this fact the contribution of the reaction in changes of the species concentration is comparable with the diffusion contributions. Moreover the ratios of the reaction and diffusion contributions are independent of time and coordinate. The dependence of the reaction rate on the initial species distribution is analyzed. It was firstly obtained that the number of the reaction zones is determined by the initial conditions and changes with time. The asymptotic long-time behaviour of the reaction rate also dependents on the initial distribution.

# I. Introduction

Behaviour of chemical, physical, biological and materials science systems can be simulated by the reaction-diffusion processes, where unlike species A and B diffuse and react [1, 2]. The key feature of many such processes is the dynamical reaction front, i.e., reaction zone [3-5]. The properties of the reaction zone are generally investigated considering two species A and B initially separated by an impenetrable barrier, which is removed at time $t = 0$ and the species diffuse and react. The behaviour of the zone for irreversible $A + B \rightarrow C$ and reversible $A + B \leftrightarrow C$ reactions was comprehensively investigated in many papers [1, 2, 5-14]. Analyses of the reversible $A \leftrightarrow B$ reaction-diffusion processes are usually based on the introduction of the effective diffusion coefficient. This allows one to reduce the task to consideration of the species diffusion only [15, 16]. The analytical solutions of this diffusion problem were obtained under several assumptions related to ratios of the species concentrations and/or of the diffusion coefficients. In [15] the solution for the infinite and semi-infinite space approximations was presented when ratios both of the coefficients and of the concentrations are large. The limit case when one species was unmovable has been considered in [16]. However, in both publications the chemical reaction was not analyzed.

In this work we consider the reversible $A \leftrightarrow B$ reaction-diffusion process when at the beginning species are mixed in the chemical equilibrium and arbitrarily distributed. At time $t = 0$ the species start to diffuse with different diffusion coefficients, disturbing the chemical equilibrium.

The reversible $A \leftrightarrow B$ reaction-diffusion processes can be used for simulation the chemical diffusion in solids [17 -19] and in liquids [20], for example.

## II. Model

Let us consider the reversible A ↔ B reaction-diffusion process in the infinite space approximation. The process is described by a solution of the following partial differential equation set:

$$\frac{\partial C_A}{\partial t} = D_A \frac{\partial^2 C_A}{\partial x^2} - Q, \tag{1}$$

$$\frac{\partial C_B}{\partial t} = D_B \frac{\partial^2 C_B}{\partial x^2} + Q, \tag{2}$$

where $Q$ is the rate of the chemical reaction given by

$$Q = k_A C_A - k_B C_B, \tag{3}$$

$C_A$ and $C_B$ are the concentrations of the species A and B, $D_A$ and $D_B$ are their diffusion constants, respectively, $k_A$ and $k_B$ are the rate constants of the reactions A→B and B→A, respectively. The initial conditions are

$$C_A(x,0) = f_A(x), \qquad C_B(x,0) = f_B(x). \tag{4}$$

Eqs. (1)-(3) and the initial conditions (4) can be represented in the following dimensionless form:

$$\varepsilon \left( \frac{\partial \tilde{C}_A}{\partial \tilde{t}} - \frac{\partial^2 \tilde{C}_A}{\partial \tilde{x}^2} \right) = -\tilde{Q}, \tag{5}$$

$$\varepsilon\left(\frac{\partial \tilde{C}_B}{\partial \tilde{t}} - D\frac{\partial^2 \tilde{C}_B}{\partial \tilde{x}^2}\right) = +\tilde{Q}, \qquad (6)$$

$$\tilde{Q} = \tilde{C}_A - k\tilde{C}_B, \qquad (7)$$

$$\tilde{C}_A(\tilde{x},0) = \tilde{f}_A(\tilde{x}), \qquad \tilde{C}_B(\tilde{x},0) = \tilde{f}_B(\tilde{x}), \qquad (8)$$

where $x = L\tilde{x}$; $t = T\tilde{t}$; $C_A = C_{A0}\tilde{C}_A$; $C_B = C_{A0}\tilde{C}_B$; $Q = k_A C_{A0}\tilde{Q}$; $\tilde{f}_A(\tilde{x}) = f_A(x)/C_{A0}$; $\tilde{f}_B(\tilde{x}) = f_B(x)/C_{A0}$; $D = D_B/D_A$; $k = k_B/k_A$; $L$, $T$ and $C_{A0}$ are the characteristic time, length, and specie concentration, respectively. We select the values of $L$ and $T$ such that $\frac{T}{L^2}D_A = 1$.

In this case the parameter $\varepsilon = (1/k_A)/T$ is the ratio of the characteristic time of the reaction $1/k_A$ to the characteristic time of diffusion $T$.

Henceforth, the symbol "~" is omitted in the notations of the dimensionless values.

### III. Solution

To find an analytical solution we will use the boundary layer function method. This method has been developed for the solution of singularly perturbed differential equations and is based on introducing a small parameter [21]. The natural parameter, which we take as a small parameter of the task, is $\varepsilon$. The condition $\varepsilon \ll 1$ corresponds to the case where the characteristic time of the reaction is much less than the characteristic time of diffusion and from (5) or (6) the chemical reaction rate $Q$ is small, $Q \sim \varepsilon \ll 1$. This means that the reaction-diffusion process is characterized by the diffusion of the species A and B which are in the chemically quasi-equilibrium state.

A solution of Eqs. (5)-(7) is seek as a sum of the regular and boundary layer parts (boundary layer functions). The regular part does not satisfy the initial conditions (8) and the purpose of the boundary layer part is to provide satisfaction of these conditions. The boundary layer functions very fast decrease with time and the regular part only describes the long-time behaviour of the reaction-diffusion process. The separation of these parts is realised by introducing different time scales: the boundary layer functions are functions of $\tau = t/\varepsilon$.

The solution of Eqs. (5)-(7) can be presented in the form of series in power of $\varepsilon$:

$$C_A(x,t,\varepsilon) = \sum_{n=0}^{\infty} \varepsilon^n C_{an}(x,t) + \sum_{n=0}^{\infty} \varepsilon^n \Pi C_{an}(x,\tau), \tag{9}$$

$$C_B(x,t,\varepsilon) = \sum_{n=0}^{\infty} \varepsilon^n C_{bn}(x,t) + \sum_{n=0}^{\infty} \varepsilon^n \Pi C_{bn}(x,\tau), \tag{10}$$

$$Q(x,t,\varepsilon) = \sum_{n=0}^{\infty} \varepsilon^n q_n(x,t) + \sum_{n=0}^{\infty} \varepsilon^n \Pi q_n(x,t). \tag{11}$$

In Eqs. (9)-(11) the first series give the regular parts and the second ones correspond to the boundary layer functions.

The initial conditions (8) are represented in the following form:

$$\sum_{n=0}^{\infty} \varepsilon^n C_{an}(x,0) + \sum_{n=0}^{\infty} \varepsilon^n \Pi C_{an}(x,0) = f_A(x), \tag{12a}$$

$$\sum_{n=0}^{\infty} \varepsilon^n C_{bn}(x,0) + \sum_{n=0}^{\infty} \varepsilon^n \Pi C_{bn}(x,0) = f_B(x). \tag{12b}$$

Substituting series (9)-(11) into Eqs. (5)-(7) and collecting coefficients at $\varepsilon^n$ for various time scales we obtain the following equations

for the regular part:

$$n = 0 \qquad q_0(x,t) = C_{a0}(x,t) - kC_{b0}(x,t) = 0, \tag{13}$$

$$n = 1,2,\ldots \qquad q_n = C_{an} - kC_{bn}, \tag{14a}$$

$$\left(\frac{\partial C_{an-1}}{\partial t} - \frac{\partial^2 C_{an-1}}{\partial x^2}\right) = -q_n, \qquad (14b)$$

$$\left(\frac{\partial C_{bn-1}}{\partial t} - D\frac{\partial^2 C_{bn-1}}{\partial x^2}\right) = q_n, \qquad (14c)$$

and for the boundary layer functions:

$n = 0$
$$\frac{\partial \Pi C_{a0}}{\partial \tau} = -\Pi q_0, \qquad (15a)$$

$$\frac{\partial \Pi C_{b0}}{\partial \tau} = \Pi q_0, \qquad (15b)$$

$n = 1, 2, \ldots$
$$\left(\frac{\partial \Pi C_{an}}{\partial \tau} - \frac{\partial^2 \Pi C_{an-1}}{\partial x^2}\right) = -\Pi q_n, \qquad (16a)$$

$$\left(\frac{\partial \Pi C_{bn}}{\partial \tau} - D\frac{\partial^2 \Pi C_{bn-1}}{\partial x^2}\right) = \Pi q_n, \qquad (16b)$$

and for all $n$ ($n = 0, 1, \ldots$) $\qquad \Pi q_n = \Pi C_{an} - k \Pi C_{bn}. \qquad (17)$

Note that except $n = 0$ all the initial conditions (12) are zero.

For $n = 0$ from Eqs. (15) taking into account that all the boundary layer functions limit to zero with time we obtain

$$\Pi C_{a0} = M_{a0}(x)\exp[-(1+k)\tau], \quad \Pi C_{b0} = -M_{a0}(x)\exp[-(1+k)\tau] \qquad (18)$$

From Eq. (13) and initial conditions (12) $C_{a0}(x,0)$, $C_{b0}(x,0)$, and $M_{a0}(x)$ are determined as:

$$C_{b0}(x,0) = \frac{1}{1+k}(f_A(x) + f_B(x)), \tag{19a}$$

$$C_{a0}(x,0) = \frac{k}{1+k}(f_A(x) + f_B(x)), \tag{19b}$$

$$M_{a0}(x) = \frac{f_A(x) - k f_B(x)}{1+k}. \tag{19c}$$

Summing (14b) and (14c) at $n = 1$ and accounting (13) we obtain homogeneous diffusion equations for $C_{a0}(x,t)$ and $C_{b0}(x,t)$, of which solutions are [22]

$$C_{a0}(x,t) = \frac{1}{\sqrt{4\pi D_{eff} t}} \int_{-\infty}^{+\infty} h(y) e^{-\frac{(x-y)^2}{4 D_{eff} t}} dy, \tag{20a}$$

$$C_{b0}(x,t) = C_{a0}(x,t)/k, \tag{20b}$$

where $D_{eff} = \frac{k+D}{1+k}$ is the effective diffusion coefficient, $h(x) = \frac{k}{1+k}(f_A(x) + f_B(x))$. From the solution one can see that the propagation of both species is determined by the same diffusion coefficient $D_{eff}$ even when one of the species is immovable, i.e. $D = 0$.

In the zeroth approximation, $n = 0$, the reaction rate is zero, $q_0 = 0$, and, usually [15, 17], the reversible A ↔ B reaction-diffusion process is considered as a diffusion with the effective diffusion coefficient $D_{eff}$ only, without analyses of the reaction. In this approximation the reaction rate is determined by only the boundary function $\Pi q_0(x,t)$ which is given by (17) and (18) and tends to zero with time according to the exponential law. At $t \gg \varepsilon$ the reaction rate is determined by the next term of the series for the reaction rate, $q_1$. This term can be determined using a zero approximation for the species concentrations $C_{a0}(x,t)$ and $C_{b0}(x,t)$. From Eq. (14b) one obtains:

$$q_1 = (1 - D_{eff}) \frac{1}{\sqrt{4\pi D_{eff} t}} \int_{-\infty}^{+\infty} h(y) \frac{d^2}{dx^2} e^{-\frac{(x-y)^2}{4 D_{eff} t}} dy. \tag{21}$$

In the framework of the assumption that the reactions A→B and B →A are fast the process is described as diffusion of species with an effective diffusion coefficient. All the information about the dominating reaction and area where reactions occur is given by the term $q_1$.

The change of the species concentration is determined by two processes: diffusion and reaction. At $t \gg \varepsilon$ the total time variations of concentrations $C_{a0}$ and $C_{b0}$ are given by $D_{eff} \partial^2 C_{a0} / \partial x^2$ and $D_{eff} \partial^2 C_{b0} / \partial x^2$, respectively, while the changes of these concentrations due to diffusion are $\partial^2 C_{a0} / \partial x^2$ and $D \partial^2 C_{b0} / \partial x^2$. The differences between these values give contribution of the chemical reaction into the concentration changes. The ratios of the reaction and diffusion contributions are independent of time and coordinate:

$$\left(\frac{\partial C_{a0}}{\partial t}\right)_{react} \bigg/ \left(\frac{\partial C_{a0}}{\partial t}\right)_{diff} = \frac{D-1}{1+k},$$

$$\left(\frac{\partial C_{b0}}{\partial t}\right)_{react} \bigg/ \left(\frac{\partial C_{b0}}{\partial t}\right)_{diff} = -\frac{(D-1)k}{D(1+k)},$$

where indexes "react" and "diff" correspond to concentration changes due to reaction and diffusion, respectively.

So, the diffusion-reaction process can be considered as a quasi-equilibrium process; however the contribution of the reaction in the concentration changes is comparable with the diffusion contribution. This is one of the main results of this paper showing that at analyses of the reversible A ↔ B reaction-diffusion process one should consider the first approximation for the reaction rate $q_1$ and of the zero approximation for the species concentrations.

In the limit case when the diffusion coefficients are equal, $D = 1$, species A and B are in the chemical equilibrium all time, $D_{eff} = 1$, and the reaction contribution is absent. In the other limit case, when one of the species is immovable, $D = 0$, the change of the species B concentration is determined by the reaction contribution only and, as result, the species B "effectively" diffuses.

The obtained zero (18) and (20) and first (21) approximations allow one to investigate the reaction-diffusion processes with arbitrary initial condition:

I. the species A and B are initially in the chemically equilibrium state ($f_A(x) - k\, f_B(x) = 0$) and this initially chemically equilibrium state is destroyed by the diffusion;

II. the species A and B are initially in the non-equilibrium state ($f_A(x) \neq k\, f_B(x)$). In framework of the considered model, establishment of the locally equilibrium state is described by the boundary layer functions and takes time of the order of $\varepsilon$ (see (18)).

To analyze the dependence of solution on the initial conditions we restrict ourselves consideration of the chemically equilibrium initial states. In this case $f_A(x) - k\, f_B(x) = 0$ and following $h(x) = f_A(x)$, $\Pi C_{a0} = 0$ and $\Pi C_{b0} = 0$. At $\varepsilon << 1$ for $t > \varepsilon$ within the accuracy of $\varepsilon$ it is enough to consider a zero approximation for the species concentration, Eqs. (20), and a first term of the series for the reaction rate, Eq. (21). The next approximations are presented in Appendix.

## IV. Influence of the initial conditions

Let us analyze influence of the initial species distribution on the reversible A ↔ B reaction-diffusion processes considering three cases.

4.1 *Stepwise initial condition*

Let us first consider the stepwise initial condition for the species concentration:

$$C_A(x,0) = \theta(-x), \quad C_B(x,0) = \theta(-x)/k, \qquad (22)$$

here $\theta(x)$ is the Heaviside step function. The task in this formulation simulates the case where the chemically equilibrium mixture of two species A and B initially occupies a half of space which is separated by an impenetrable barrier form an empty space. The barrier is removed at time $t = 0$ and the species diffuse with different diffusion coefficients and react. For example, this task appears at analysis of the interdiffusion and the "Kirkendall effect" in the solids [23]. From (20a) we obtain

$$C_{a0s} = \frac{1}{2}\mathrm{erfc}(\xi) \qquad (23)$$

and from (21) the reaction rate is

$$q_{1s} = \frac{\alpha}{t}\frac{e^{-\xi^2}}{\sqrt{\pi}}\xi, \qquad (24)$$

where $\alpha = \dfrac{(1-D)}{2(D+k)}$, $\xi = \dfrac{x}{2\sqrt{D_{\mathit{eff}}\, t}}$.

The reaction rate as a function of $\xi$ is presented in Fig. 1. There are two reaction zones at $x > 0$ and $x < 0$. Note that the reaction rate sign (the dominating reaction) is determined by the ratio of the diffusion coefficients, $D$. If $D < 1$ reaction A→B dominates at $x > 0$ and reaction B

→A prevails at $x < 0$. At $D > 1$ the reversed situation is observed. This principally distinguishes our results from the results for the case of the irreversible A + B → C diffusion-reaction with the initial separated species, where there is a single reaction zone [5]. Under special conditions the alternating-sign reactions have been obtained in [24, 25], where the reversible A + B ↔ C reaction-diffusion processes were considered. As distinct from [24, 25] the reversible A ↔ B reaction-diffusion process considered here the alternating-sign reaction is the general result obtained without any additional assumptions.

We can characterize propagation of the reaction zone as propagation of the maximum reaction rate, the coordinate $x_{max}$ of which increases with time proportionally to $t^{1/2}$. This propagation is determined by the effective diffusion coefficient $D_{eff}$. The width of the reaction zone can be introduced as a half-width which increases $\sim t^{1/2}$ and equals about $x_{max}$. The maximum of the reaction rate and total rate, determined as integral through a half of space, decrease with time as $1/t$ and $1/t^{1/2}$, respectively. The dependences of the coordinate $x_{max}$, reaction zone width, and reaction rates on time are in accord with the asymptotic long-time approximation for the reversible A + B ↔ C reaction-diffusion processes [11, 12], where a small parameter was $1/t$. In difference from [11,12] in our model these dependences are valid at all the time because of the small parameter selected by us is the ratio of the characteristic time of the reaction to the characteristic time of diffusion.

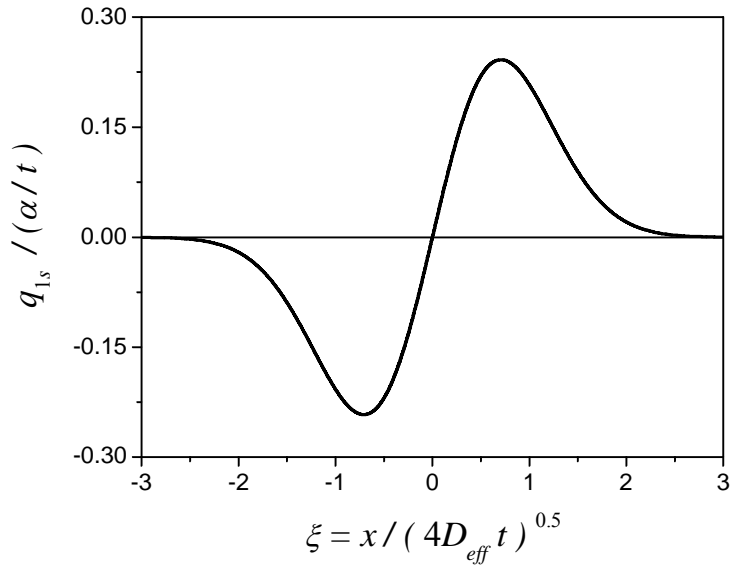

Fig. 1. The normalized reaction rate as a function of $\xi$.

*4.2 Island of A and B species surrounded by an empty space*

Recently, a new line in the study of the irreversible A + B → 0 reaction-diffusion processes has been developed under the assumption that the particle number of one or both species is finite [13, 14]. The consideration of the finite space tasks gives qualitatively new results. For example, it was shown [14] that the process can be divided into two stages in time. At time much less than the characteristic diffusion time the front characteristics are described by the well-known power law dependencies on time, whereas these are well approximated by exponential laws in the opposite case. At the time when the power laws change by the exponential ones, the total quantities of the reactants are about 0.5 of their initial quantities. As the second example let us consider an island of A and B initially in the equilibrium state surrounded by an empty space. The initial conditions are the following:

$$C_A(x,0) = \begin{cases} 0 & \text{if } |x| > a \\ 1 & \text{if } |x| \leq a \end{cases} \quad \text{and} \quad C_B(x,0) = C_A(x,0)/k. \tag{25}$$

The solution for the concentration $C_A$ can be obtained as a difference of two solutions (24) shifted by $-a$ and $a$:

$$C_{a0p} = \frac{1}{2}\left[\text{erfc}\left(\frac{x-a}{\sqrt{4D_{eff}t}}\right) - \text{erfc}\left(\frac{x+a}{\sqrt{4D_{eff}t}}\right)\right]. \tag{26}$$

Using Eq. (24) the reaction rate can be presented as

$$q_{1p} = \frac{\alpha}{t\sqrt{4\pi D_{eff}t}}\left[(x-a)\exp\left(-\frac{(x-a)^2}{4D_{eff}t}\right) - (x+a)\exp\left(-\frac{(x+a)^2}{4D_{eff}t}\right)\right]. \tag{27}$$

Analogically to the first example a type of the dominating reaction is determined by the value of the dimensionless diffusion coefficient $D$. The process can be separated into two stages. During the first stage the reaction at $t < a^2/4D_{eff}$ four reaction zones are observed: two zones are near every boundary of the island (Fig. 2). With the time the zones inside the island merge into single and during the second stage there are three reaction zones. The reaction rate $q_1$ in the island centre is about zero in the initial stage of the process, at $t \ll a^2/4D_{eff}$, and increases with time during the first stage (Fig. 3). In this time all the maximums of modules of the reaction rate decrease (Fig. 2).

At the asymptotic long-time approximation, $t \gg a^2/4D_{eff}$, solution (27) is reduced to

$$q_{1p}/(\xi_0 \alpha/t) \approx \frac{2}{\sqrt{\pi}}\left(2\xi^2 - 1\right)e^{-\xi^2},$$

which is presented in Fig. 4 (here $\xi_0 = a/\sqrt{4D_{eff}t}$ ). All the extremums decrease with time as $1/t^{3/2}$.

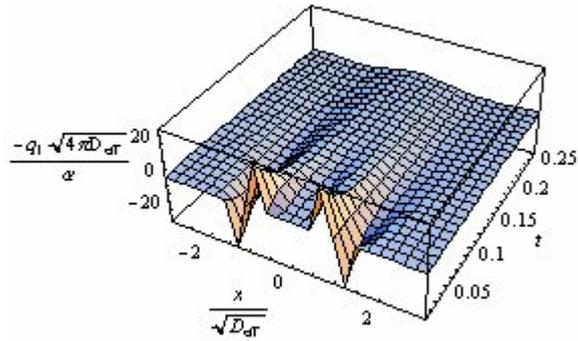

Fig. 2 The normalized reaction rate as a function of $x$ and $t$ for $a/\sqrt{D_{eff}} =1$. Note that for clearness this plot presents $-q_1$.

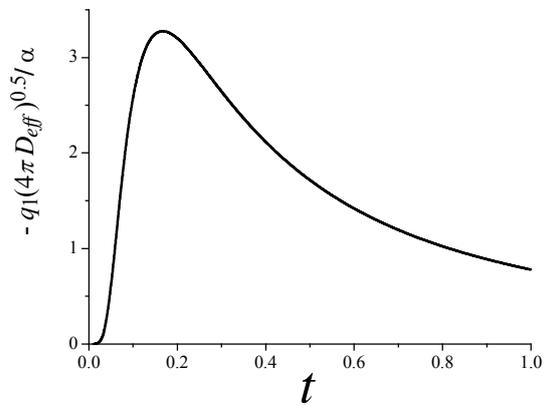

Fig. 3. The module of the normalized reaction rate in the island centre ($x = 0$) as a function of time.

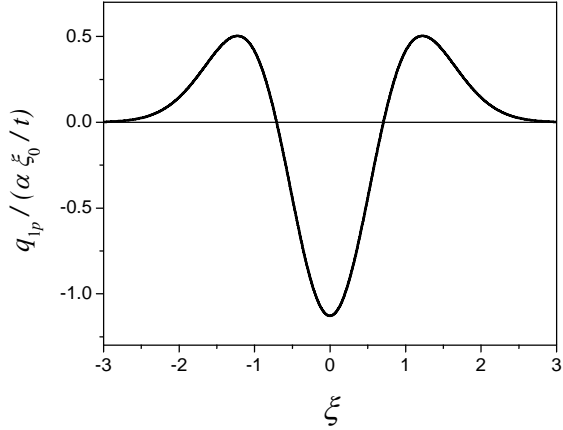

Fig. 4. The asymptotic long-time approximation for the reaction rate $q_{1p}$.

4.3 *Perturbation of the chemically equilibrium state*

As the third example let us consider an alternative-sign perturbation of the chemically equilibrium state. We simulate the perturbation by the following initial conditions:

$$C_A(x,0) = \begin{cases} C_0 & \text{if} \quad |x| > a \\ C_0 + 1 & \text{if} \quad -a \leq x \leq 0 \\ C_0 - 1 & \text{if} \quad 0 < x \leq a \end{cases} \quad \text{and} \quad C_B(x,0) = C_A(x,0)/k. \qquad (28)$$

The solution for the concentration $C_A$ can be obtained as a difference of two solutions (26) changed $a$ by $a/2$ and shifted by $-a/2$ and $a/2$:

$$C_{a0d} = C_0 + \frac{1}{2}\left[-\text{erfc}\left(\frac{x-a}{\sqrt{4D_{eff}t}}\right) + 2\text{erfc}\left(\frac{x}{\sqrt{4D_{eff}t}}\right) - \text{erfc}\left(\frac{x+a}{\sqrt{4D_{eff}t}}\right)\right]. \qquad (29)$$

Using Eq. (29) the reaction rate (Fig. 5) is presented as

$$q_{1d} = \frac{\alpha}{t\sqrt{4\pi D_{eff} t}} \left[ 2x\exp\left(-\frac{x^2}{4D_{eff} t}\right) - (x-a)\exp\left(-\frac{(x-a)^2}{4D_{eff} t}\right) - (x+a)\exp\left(-\frac{(x+a)^2}{4D_{eff} t}\right) \right]. \quad (30)$$

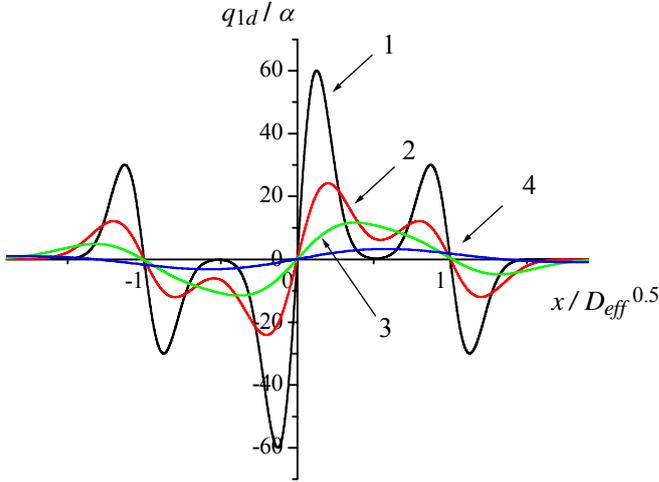

Fig. 5. Reaction rate for the case of perturbation of the chemically equilibrium state: $1 - t = 0.008$; $2 - t = 0.02$; $1 - t = 0.05$; $1 - t = 0.2$.

In the initial stage of the process there are 6 extremums, two extremums at every step boundary of the initial conditions. Inside the area of the perturbation four extremums run into two ones with time (lines 3 and 4, Fig. 5). At asymptotic long-time approximation, $t \gg a^2/4D_{eff}$, solution (27) is reduced to the following self-similar form:

$$q_{1d} / \left(\xi_0^2 \alpha / t\right) \approx \frac{2\xi}{\sqrt{\pi}} \left(3 - 2\xi^2\right) e^{-\xi^2}.$$

The extremums of the reaction rate decrease with a time increase as $1/t^2$.

## V. Conclusion

The investigation of the reversible A ↔ B reaction-diffusion process, when species A and B are initially mixed and diffuse with different diffusion coefficients using the boundary layer function method shows that the process can be considered as a quasi-equilibrium process. Despite this fact the contribution of the reaction in changes of the species concentration is comparable with the diffusion contributions. Moreover the ratios of the reaction and diffusion contributions are independent of time and coordinate.

The ratio of the characteristic time of the reaction to the characteristic time of diffusion is considered as a small parameter of the task. This approach allowed us to consider influence of the initial distribution of species on the process and to obtain analytical solutions for model problems at all the time. It was firstly obtained that the number of the reaction zones is determined by the initial conditions and changes with time. The asymptotic long-time behaviour of the reaction rate also dependents on the initial distribution.

The analysis of the higher number terms shows that their contribution not only is proportional to the small parameter but also decreases with time.

**Appendix**

To analyze area of applicability of the approximations obtained above let us find the next terms of the series for the case where the species A and B are initially in the chemically equilibrium state ($f_A(x) - k f_B(x) = 0$). The solution of Eqs. (16) - (17) for boundary layer parts at $n = 1$ are

$$\Pi C_{a1} = M_{a1} e^{-(1+k)\tau}, \qquad \Pi C_{b1} = -M_{a1} e^{-(1+k)\tau}, \qquad (A.1)$$

where $M_{a1}$ is the function of coordinate $x$ determined from Eqs. (12) at $n = 1$.

From Eqs. (12) and (A.1) it is following that $C_{a1}(x,0) = -C_{b1}(x,0)$ and a solution of Eqs. (14) at $n = 2$ is given by

$$C_{a1}(x,t) = \frac{q_1(x,t)}{1+k} + \frac{t}{1+k}(D_{eff} - D)\frac{\partial^2 q_1}{\partial x^2},$$ (A. 2)

$$C_{b1}(x,t) = -\frac{q_1(x,t)}{1+k} - \frac{t}{1+k}(D_{eff} - 1)\frac{\partial^2 q_1}{\partial x^2}.$$ (A. 3)

and from Eq. (12) function $M_{a1} = M_{a1}(x)$ is obtained as:

$$M_{a1} = \frac{(1 - D_{eff})}{1+k}\frac{d^2 h(x)}{dx^2}$$ (A.4)

Now from Eq. (14b) the next approximation for the reaction rate is

$$q_2 = \frac{(1 + D - 2D_{eff})}{1+k}\frac{\partial^2 q_1}{\partial x^2} + \frac{t(D_{eff} - D)}{1+k}\frac{\partial^4 q_1}{\partial x^4}.$$ (A. 5)

In contrast to the zero approximation we obtained non-zero boundary layer parts (see Eqs. (A.1) and (A.4)) and the solution is given by the regular parts only at $t \gg \varepsilon$, when the boundary layer parts is vanishingly small. From the structure of the obtained expressions, Eqs. (24), (27), and (30), for $q_1$, one can see that at $t \gg \varepsilon$ the next terms ($n = 1$) of the series for the concentrations are presented as bounded functions multiplied by $\varepsilon/t$ and $q_2 \sim (\varepsilon/t)^2$. Such, the small parameter is $\varepsilon/t$ and significance of the higher number terms decreases with time.